\documentstyle[11pt]{article}
\begin{document}
\pagestyle {headings}
\thispagestyle{plain}
\pagenumbering{arabic}
\newcommand{\gap}{\lower .7ex\hbox{$\;\stackrel{\textstyle >}{\sim}\;$}}
\newcommand{\lap}{\lower .7ex\hbox{$\;\stackrel{\textstyle <}{\sim}\;$}}
\parskip=1ex
{\large
\hfill{
\begin{tabular}{l}
DSF-T-92/23\\
INFN-NA-IV-92/23
\end{tabular}}}

\bigskip

\LARGE
\centerline {{\bf Nonleptonic Cabibbo favoured ${\bf B}$-decays}}
\bigskip
\centerline  {{\bf and ${\bf CP}$-asymmetries for Charmed final hadron}}
\bigskip
\centerline {{\bf states in Isgur and Wise theory}\footnote{To be published
in {\it Zeitschrift f\"ur Physik C.}} }
\Large
\vspace{2.0cm}
\centerline { F.\ Buccella $^{1}$, F.\ Lombardi $^{1}$,
G.\ Miele $^{1,2}$ and P.\ Santorelli $^{1,2}$}
\normalsize
\bigskip\bigskip
\par\noindent
$^{1}$ Dipartimento di Scienze Fisiche ,Universit\`a di Napoli, Mostra
 D'Oltremare pad.19, I--80125 Napoli, Italy\\
\\
$^{2}$INFN, Sezione di Napoli,
Mostra D'Oltremare pad.20, I--80125 Napoli, Italy

\vspace{.5cm}

\begin{center}
\end{center}

\vspace{.5cm}

\par\noindent
\baselineskip=0.5cm
\centerline{{\bf Abstract}}

\bigskip

The Cabibbo allowed non-leptonic $B$-decays in two hadrons are studied,
within the factorization hypothesis, in the framework of Isgur and Wise
theory for the matrix elements of the $\Delta B=-\Delta C=\pm 1$ weak
currents. The $SU(2)_{HF}$ symmetry relates $|\Delta B|=1$ to
$|\Delta C|=1$ currents, which have  been measured in the semileptonic
strange decays of charmed particles. By assuming colour screening
and allowing for $SU(3)$ invariant contributions  from the annihilation
terms with charmed final states one is able to comply with the present
experimental knowledge.\\
The $CP$ violating asymmetries in neutral $B$ decays are given for
charmed final states in terms of the $K-M$ angles. With the central
values found for the annihilation parameters there is a destructive
(constructive) interference between the direct and annihilation terms
in the Cabibbo allowed (doubly forbidden) amplitudes for the decays
into $D^{0}(D^{*0})\pi^0$ and $D^0\rho^0$ so that they may be of the same
order. This would imply large asymmetries, for which however our
present knowledge on the amplitudes does not allow to predict even
their sign.\\
We have better confidence in our predictions for the charged final
states than the neutral ones and can draw the conclusion that the
detection of the corresponding asymmetries requires, at least,
$10^6$ tagged neutral $B$-particles.
\newpage
\baselineskip=0.7cm
\bigskip\bigskip
\par\noindent
{\bf  Introduction}
\bigskip
\par
The study of the semileptonic decays of $B$ particles in final hadrons
with $|C|=1,0$ provides a laboratory to test Isgur and Wise theory [1].\
Following Neubert [2] we can get for $|V_{bc}|$,
by an extrapolation of the
data [3] on the semileptonic decay  $\overline{B}_d^0\rightarrow D^{+*} e^{-}
\bar\nu_e$, the range of values $0.044\div 0.052$ in agreement with
previous determinations.\\
The smallness of the element $V_{bu}$ does not allow, at the moment, to
have information about $\Delta B=1$ and $\Delta C=0$ matrix elements of
the vector and axial currents.
A larger domain to compare theory with the
experiment would be obtained in the framework of the factorization
hypothesis since the matrix
elements of the $\Delta B=1$ weak currents appear in the amplitudes of
the exclusive two-body nonleptonic
decays. In a recent paper [4] about nonleptonic decays of charmed particles
one has been able to find a reasonable fit to the amplitudes for $PP$, $PV$
and $PA$
final states by keeping into account the final state interaction and by
taking as a free parameters the matrix elements of the divergences of the
vector and axial currents between the vacuum and two hadrons final states,
which appear in the annihilation terms.
For the amplitudes of $B$ decays we neglect the final state interaction
of the two hadrons, due to the high energy of the decay products,
and, as for charmed decays,
we take as free parameters, related by $SU(3)$ symmetry,
the matrix elements of the divergences of the weak currents between
the vacuum and the two mesons $|C|=1$ final states.

The weak amplitudes, within the factorization approximation, depend on
the matrix elements of the weak currents between $B=1$ and $C=1$
mesons, which may be written in terms of the universal Isgur-Wise
function $\xi(w^2)~$[1], or mesons built with light quarks ($u$, $d$ and
$s$).\\
The $E691$ Collaboration [5] has studied the lepton spectrum in the
semileptonic
decays of charmed particles in $S=1$ final hadron state and
found the residua of the relevant form factors within the
hypothesis of pole dominance and with the position of the pole dictated
by the lowest resonance with the right quantum numbers. The matrix elements
between $B=1$ and the light mesons may be related to the ones, given in [5],
according to $SU(2)_{HF}$ [6],[7].
In conclusion we are able to predict the amplitudes for Cabibbo favoured decays
of $B$ mesons, for which one has some experimental knowledge, in terms
of the Isgur-Wise function
$\xi(w^2)$ of the three parameters appearing in the annihilation
terms and in terms of the $|\Delta C|=1$ form factors, which have been
measured by $E691$ Collaboration. We shall take  the parabolic expression
previously proposed [8] for $\xi(w^2)= 1+bw^2+cw^4$ as well
as different
choices [9] to show the dependence of the results on the parameterization.

In previous works [10] we studied the $CP$ violating asymmetries expected in
the decays of the neutral $B$ particles still in the framework of the
factorization approximation and neglecting the final state interaction,
but with different assumptions on the matrix
elements of the weak currents. Within the theoretical framework described
here we are able to predict the asymmetries
for $|C|=1$ final states and  compare them with our previous papers.

The paper is organized as follows. In section 1 we give the effective
hamiltonian for the Cabibbo favoured nonleptonic decays of $|B|=1$
particles, the relevant matrix elements of the $\Delta B = - \Delta C=\pm 1$
form factors
vector and axial current in the Isgur and Wise theory, in terms of  $\xi(w^2)$
and the relationships between $\Delta B = \pm 1$ and $\Delta C = \pm 1$
form factors deduced from $SU(2)_{HF}$ in the framework
Heavy Quark Effective Theory (HQET) [6],[11].

In the second section we compute some of the considered amplitudes
and we define the free parameters appearing in the annihilation terms.
For $\xi(w^2)$ we assume the parabolic, exponential and pole
form respectively
(with $\xi(0)=1$), whereas the form factors, which appear in $\Delta B = 1$
currents, will be written in terms of the experimentally fitted for
$\Delta C=1$. Furthermore, releasing the $\xi(0)=1$ constraint (dictated
by $HQET$) we shall also study the agreement with the experimental data
of the theoretical predictions obtained assuming for $\xi(w^2)$ a physical
pole expression.

In the third section we shall give our predictions for the $CP$ violating
asymmetries for $C=\pm 1$ final states and show their dependence on the
parameters appearing in the annihilation terms.

Finally we give our conclusions.

\par\noindent
{\bf 1. Matrix elements of $\bf{\Delta B= \pm 1 }$ weak currents in HQET}
\bigskip
\par
The bare weak Hamiltonian for the $|\Delta B|=1$ Cabibbo favoured
nonleptonic decays is:
\begin{equation}
 H^{\Delta B=\pm 1} =-i\frac{4G_F}{\sqrt{2}}V_{bc}
({\bar c_L}\gamma_{\mu}b_L)\left(V^*_{sc}{\bar s_L}\gamma^{\mu}c_L+
V^*_{du}{\bar d_L}\gamma^{\mu}u_L\right )+ h.c.~~~.
\label{1}
\end{equation}
The short-range $QCD$ corrections at  subleading -log approximation have been
computed by the rinormalization group approach [12].\\
We can get $H^{\Delta B=\pm 1}_{eff}$ from (\ref{1}) by performing the
following substitution
\footnote
{We neglect the penguin contribution to the $\Delta B=\Delta S=\pm 1$
amplitudes   since the small coefficient $\approx 0.028$ for the corresponding
operator is not compensated, as in the decays of strange particles, by an
enhancement for the matrix elements, since the final hadrons do not contain
quarks of the first family.}
\begin{eqnarray}
{\bar c_L}\gamma_{\mu}b_L~{\bar q_{2L}}\gamma^{\mu}q_{1L}
&\rightarrow & k_1~{\bar c_L}\gamma_{\mu}b_L~{\bar
q_{2L}}\gamma^{\mu}q_{1L}
\nonumber \\
& + & k_2~{\bar q_{2L}}\gamma_{\mu}b_L~{\bar c_{L}}\gamma^{\mu}q_{1L}
\label{2}
\end{eqnarray}
with $q_1 \equiv c~$(or $u$), $q_2 \equiv s~$(or $d)$ and $k_1$, $k_2$
given by:
\begin{eqnarray}
C_{+} & = & \left(\frac{\alpha_s(m_b)}{\alpha_s(m_W)}\right)^{-\frac{6}{23}}
\left(1-0.51\frac{\alpha_s(m_b)-\alpha_s(m_W)}{\pi}\right)
\nonumber\\
C_{-} & = & \left(\frac{\alpha_s(m_b)}{\alpha_s(m_W)}\right)^{\frac{12}{23}}
\left(1+1.48\frac{\alpha_s(m_b)-\alpha_s(m_W)}{\pi}\right)
\nonumber\\
k_1 & \equiv & \frac{1}{2}\left(C_+ + C_- \right)\cong 1.11
\nonumber\\
k_2 & \equiv & \frac{1}{2}\left(C_+ - C_- \right)\cong -0.25
\label{3}
\end{eqnarray}
The values for $\alpha_s(m_{b})$ and $\alpha_s(M_{W})$ are taken
consistently with the value for $\alpha_s(m_{c})$ of a previous work
on the nonleptonic decays of $D$ particles [4]:
\begin{equation}
\alpha_{s}(m_{c})=0.272
 ~~~~~~~~~~~ \alpha_{s}(m_{b})= 0.187
{}~~~~~~~~~~~ \alpha_{s}(M_{W}) =0.110
\label{4}
\end{equation}
In the factorization approximation the matrix elements of
$H^{\Delta B = \pm 1}_{eff}$ are given in terms of the matrix elements of
the vector and axial currents as:
\begin{eqnarray}
\langle D^{0}|\bar{c} \gamma_{\mu} b | B^{-} \rangle &  &
\langle D^{*0}|\bar{c} \gamma_{\mu} \gamma_{5}b | B^{-} \rangle
\nonumber\\
\langle K^{-}|\bar{s} \gamma_{\mu} b | B^{-} \rangle & &
\langle \rho^{-}|\bar{d} \gamma_{\mu} \gamma_{5}b | B^{-} \rangle
\nonumber\\
\langle D^{-} \bar{K}^{0}| \partial^{\mu} \bar{s} \gamma_{\mu} c | 0 \rangle
& &
\langle D^{*-} \bar{K}^{0}| \partial^{\mu} \bar{s} \gamma_{\mu} \gamma_{5}c |0
\rangle
\nonumber\\
& \langle 0|\bar{d} \gamma_{\mu} \gamma_{5} b | \bar{B}^{0}_{d} \rangle &
\label{5}
\end{eqnarray}
The matrix elements between charmed and beautiful mesons are given
in terms of the universal form factor $\xi(w^2)$ according
to Isgur and Wise theory [11]:
\begin{eqnarray}
& &\langle
D(v')|\bar{c}\gamma_{\mu} b|B(v)\rangle=C_{bc}\sqrt{m_Bm_D}\xi (w^2)
(v_{\mu}+v'_{\mu})\\
& & \langle D^*(v',\varepsilon )|\bar{c}\gamma_{\mu} \gamma_5 b|B(v)\rangle=
i C_{bc}\sqrt{m_Bm_{D^*}}
\xi (w^2)\left [\varepsilon_{\mu}^*(1+v\cdot v')-(\varepsilon^*
\cdot v) v'_{\mu}\right]
\label{6}
\end{eqnarray}
where $v$ and $v'$ are the quadri-velocities of the heavy hadrons and $C_{bc}$,
neglecting the dependence on $v \cdot v'$, is given by [1],[11]:
\begin{equation}
C_{bc}=\left (\frac {\alpha_s(m_b)}{\alpha_s(m_c)}\right)^{-6/25}\cong 1.10
\label{7}
\end{equation}
The matrix elements between the $B=1$ and the light mesons $q \in \{u,d,s\}$
may be written in the general form:
\begin{eqnarray}
\langle P_{i}(p_i)| \bar{q} \gamma^{\mu} b | B(v) \rangle
& = & f_{+}^{(bq)} (q^2)
( m_{B}~ v^{\mu} +  p_{i}^{\mu} ) + f_{-}^{(bq)} (q^2) (m_{B}~v^{\mu}-
p_{i}^{\mu})
\nonumber\\
\langle V_{j}(p_j,\varepsilon) | \bar{q} \gamma^{\mu}\gamma_5 b | B(v) \rangle
 &  = &  i f^{(bq)} (q^2)\varepsilon^{*\mu}+
i a_{+}^{(bq)} (q^2)\varepsilon^*\cdot v~ m_{B}( m_{B}~ v^{\mu} +  p_{j}^{\mu}
)
\nonumber\\
& + &
i a_{-}^{(bq)} (q^2)\varepsilon^*\cdot v~ m_{B}( m_{B}~
v^{\mu} -  p_{j}^{\mu} )
\label{8}
\end{eqnarray}
and in the same way for $C=1$:
\begin{eqnarray}
\langle P_{i}(p_i) | \bar{q} \gamma^{\mu} c | D(v) \rangle & = & f_{+}^{(cq)}
(q^2)
( m_{D}~ v^{\mu} +  p_{i}^{\mu} ) + f_{-}^{(cq)} (q^2) (m_{D}~v^{\mu}-
p_{i}^{\mu})
\nonumber\\
\langle V_{j}(p_j,\varepsilon) | \bar{q} \gamma^{\mu}\gamma_5 c | D(v) \rangle
&  = & i f^{(cq)} (q^2)\varepsilon^{*\mu}+
i a_{+}^{(cq)} (q^2)\varepsilon^*\cdot v~ m_{D}( m_{D}~ v^{\mu} +  p_{j}^{\mu}
)
\nonumber\\
& + &
i a_{-}^{(cq)} (q^2)\varepsilon^*\cdot v~ m_{D}
( m_{D}~ v^{\mu} -  p_{j}^{\mu} ).
\label{8bis}
\end{eqnarray}
The symmetry $SU(2)_{HF}$ allows to relate the form factors introduced in
(\ref{8}) and (\ref{8bis}) [6]:
\begin{eqnarray}
f_{+}^{(bq)}(q_{B}^{2})& = &\frac{C_{bc}}{2\sqrt{m_{B}m_{D}}}
\left[(m_{B}+m_{D})f_{+}^{(cq)}(q_D^2)+(m_{D}-m_{B})f_{-}^{(cq)}(q_D^2)\right]
\nonumber\\
\nonumber\\
f_{-}^{(bq)}(q_{B}^{2})& = &\frac{C_{bc}}{2\sqrt{m_{B}m_{D}}}
\left[(m_{D}-m_{B})f_{+}^{(cq)}(q_D^2)+(m_{B}+m_{D})f_{-}^{(cq)}(q_D^2)\right]
\nonumber\\
\nonumber\\
a_{+}^{(bq)}(q_{B}^{2})& = &\frac{C_{bc}}{2}\sqrt{{m_{D}\over m_{B}}}
\left[\left( {m_{D}\over m_{B}}+1\right)a_{+}^{(cq)}(q_D^2)
+\left( {m_{D}\over m_{B}}-1\right)a_{-}^{(cq)}(q_D^2) \right]
\nonumber\\
\nonumber\\
a_{-}^{(bq)}(q_{B}^{2})& = &\frac{C_{bc}}{2}\sqrt{{m_{D}\over m_{B}}}
\left[\left( {m_{D}\over m_{B}}-1\right)a_{+}^{(cq)}(q_D^2)
+\left( {m_{D}\over m_{B}}+1\right)a_{-}^{(cq)}(q_D^2) \right]
\nonumber\\
\nonumber\\
f^{(bq)}(q^2_{B})& = & C_{bc}\sqrt{{m_{B}\over m_{D}}}f^{(cq)}(q^2_{D})
\label{9}
\end{eqnarray}
where $C_{bc}$  is given by (\ref{7}) and
\begin{eqnarray}
q^2_{B}& = & (m_{B}~v-p_{i(j)})^2
\nonumber \\
q^2_{D}& = & (m_{D}~v-p_{i(j)})^2
\label{10}
\end{eqnarray}
The form factors for charmed and beautiful particles in
(\ref{8})and (\ref{8bis}) are related to the ones defined in the helicity
frame of reference (see for example [13]) by:
\begin{eqnarray}
f_{+}^{(Qq)}(q^2) & = & f_{1}^{(Qq)}(q^2)
\nonumber\\
f_{-}^{(Qq)}(q^2) & = & \frac{m_1^2-m_2^2}{q^2}\left[ f_{0}^{(Qq)}(q^2)-
f_{1}^{(Qq)}(q^2)\right]
\nonumber\\
f^{(Qq)}(q^2) & = &(m_1+m_2) A_{1}^{(Qq)}(q^2)
\nonumber\\
a_{+}^{(Qq)}(q^2) & = &-\frac{A_{2}^{(Qq)}(q^2)}{m_1+m_2}
\nonumber\\
a_{-}^{(Qq)}(q^2) & = &-\frac{1}{q^2}\left[ (m_1+m_2)A_{1}^{(Qq)}(q^2)-
(m_1-m_2)A_{2}^{(Qq)}(q^2)+2m_2A_{0}^{(Qq)}(q^2)\right]\nonumber\\
\label{11}
\end{eqnarray}
where $m_1$ and $m_2$ stand for the masses of the initial and final particle
respectively; whereas $Q \in \{b,c\}$ and $q \in \{u,d,s\}$.\\
The experimental data for the semileptonic decays of $D$ particles
with $S=-1$ in the final state, allow to fit the form factors defined
in the r.h.s.'s of (\ref{11}) in the case of $Q=c$ and $q=s$,
by assuming for them a pole behaviour dominated by
the lower resonance with the proper quantum numbers and leaving the
residua as free parameters [5]. We extend these results to
$q=d(u)$, applying $SU(3)$ symmetry to the above
residua and using the correct masses for the involved resonances.\\
The results found are reported in Table 1.

\bigskip\bigskip
\par\noindent
{\bf 2. Evaluation of the amplitudes and comparison with experiment}
\bigskip
\par
We write the amplitudes for the nonleptonic decays of $B$ particles for
some particular processes  assuming a  complete colour screening
\footnote{We extend to the $B$ particles nonleptonic decays, the results of
$D$ mesons decay [4], which suggest a complete screening of the
factorizable part of the amplitude proportional to $1/N_c$ due to the
non-factorizable one [14].}:
\begin{eqnarray}
{\cal A}(B^{-} \rightarrow D^{0} \pi^{-}) & = & i {G_{F} \over \sqrt{2} }
V_{bc} V_{du}^{*} \left[ k_{1} \langle D^{0} |  \bar{c} \gamma_{\mu} b | B^{-}
 \rangle
\langle \pi^{-} | \bar{d} \gamma^{\mu} \gamma_{5} u | 0  \rangle
\right.
\nonumber\\
& + &\left. k_{2}\langle \pi^{-}|{\bar d}\gamma_{\mu} b|B^{-}\rangle
\langle D^{0}|{\bar c}\gamma_{\mu} \gamma_{5} u|0\rangle\right ]
\nonumber\\
& = &  {G_{F} \over \sqrt{2} } V_{bc} V_{du}^{*}\left [ k_{1} C_{bc}
\sqrt{m_{B}
m_{D}} \xi(w_{\pi}^{2}) f_{\pi}
\left(m_{B}- m_{D} \right) \left(1 + {E_{D} \over m_{D} } \right)
\right.
\nonumber\\
& + & \left. k_{2} f_{0}^{(bd)}(m_{D}^2)f_{D}\left(m_{B}^2-m_{\pi}^2\right)
\right]
\label{12}
\end{eqnarray}
\begin{eqnarray}
{\cal A}(\overline{B}^{0}_{d} \rightarrow D^{+} \pi^{-}) & = & i {G_{F} \over
\sqrt{2} } V_{bc} V_{du}^{*} \left[ k_{1} \langle D^{+} |  \bar{c} \gamma_{\mu}
 b | \overline{B}^{0}_{d} \rangle \langle \pi^{-} | \bar{d}
\gamma^{\mu} \gamma_{5}  u | 0  \rangle \right.
\nonumber\\
& + &
\left. k_{2} \langle D^{+} \pi^{-} |  \bar{c} \gamma_{\mu}
 u | 0 \rangle \langle 0 | \bar{d} \gamma^{\mu} \gamma_{5}
 b | \overline{B}^{0}_{d}   \rangle \right]
\nonumber\\
& = & {G_{F} \over \sqrt{2} } V_{bc} V_{du}^{*} \left[ k_{1} C_{bc}
\sqrt{m_{B} m_{D}} \xi(w_{\pi}^{2}) f_{\pi}
\left(m_{B}- m_{D} \right) \left(1 + {E_{D} \over
m_{D} } \right) \right.
\nonumber\\
& - & \left. k_{2} (m_{c}-m_{u}) W_{PP} m_{B}^{2} \right]
\label{13}
\end{eqnarray}
\begin{eqnarray}
{\cal A}(\overline{B}^{0}_{d} \rightarrow D^{0} \rho^{0}) & = &- i {G_{F} \over
\sqrt{2} } V_{bc} V_{du}^{*}k_{2} \left[
\langle \rho^{0} |  \bar{d} \gamma_{\mu}\gamma_{5}
 b | \overline{B}^{0}_{d} \rangle \langle D^{0} | \bar{c}
\gamma^{\mu}\gamma_{5} u | 0  \rangle \right.
\nonumber\\
& + &
\left. \langle D^{0} \rho^{0} |  \bar{c} \gamma_{\mu}\gamma_{5}
 u | 0 \rangle \langle 0 | \bar{d} \gamma^{\mu}\gamma_{5}
 b | \overline{B}^{0}_{d}   \rangle \right]
\nonumber\\
& = & -i {G_{F} \over \sqrt{2} } V_{bc} V_{du}^{*}k_{2}
\left(-{1 \over \sqrt{2}}\right)
\left[  A_{0}^{(bd)}(m_{D}^2) f_{D} \right.
\nonumber\\
& - & \left. (m_{c}+m_{u}) W_{DV} \right] 2 m_{B} m_{\rho}
\varepsilon \cdot v
\label{14}
\end{eqnarray}
\begin{eqnarray}
{\cal A}(\overline{B}^{0}_{d} \rightarrow D^{+*} \pi^{-}) & = &- i {G_{F} \over
\sqrt{2} } V_{bc} V_{du}^{*} \left[ k_{1} \langle D^{+*} |  \bar{c}
\gamma_{\mu}
\gamma_{5} b | \overline{B}^{0}_{d} \rangle \langle \pi^{-} | \bar{d}
\gamma^{\mu}\gamma_{5} u | 0  \rangle \right.
\nonumber\\
& + &
\left. k_{2} \langle D^{+*} \pi^{-} |  \bar{c} \gamma_{\mu}\gamma_{5}
 u | 0 \rangle \langle 0 | \bar{d} \gamma^{\mu} \gamma_{5}
 b | \overline{B}^{0}_{d}   \rangle \right]
\nonumber\\
& = & -i {G_{F} \over \sqrt{2} } V_{bc} V_{du}^{*}
\left[ k_{1} C_{bc} \left( {{ m_{B}+m_{D^*}}\over{2 \sqrt{m_{B} m_{D^*}}}}
\right) \xi(w_{\pi}^{2}) f_{\pi} \right.
\nonumber\\
& - & \left. k_{2} (m_{c}+m_{u}) W_{D^{*}P} \begin{array}{c} \\ \end{array}
 \right] 2 m_{B} m_{D^{*}} \varepsilon \cdot v
\label{15}
\end{eqnarray}
In (\ref{12})-(\ref{15}) we denote with $W_{PP}$, $W_{DV}$ and
$W_{D^{*}P}$ the reduced matrix elements for the annihilation terms defined
analogously as in [4].
By applying $SU(3)$ symmetry to the scalar and pseudoscalar densities
one gets the corresponding contributions for the strange final states
as $D^+_s K^-$. Note that the $SU(4)$ flavour symmetry would imply
$W_{DV}=W_{D^*P}$
\footnote{The decay constants used are: $f_{D^*}=f_{D_s}=f_D=210~MeV$
(Mannel et al. in [9]), $f_{J/\Psi}=382~MeV$ [14],
$f_{\rho}=221~MeV$, $f_{K^*}=f_{K}=170~MeV~$ and $f_{\pi}=132~MeV~$.}.

{}From the experimental data on the exclusive branching ratios of the
nonleptonic decays of $B$ particle [15], the rates of the semileptonic channel
$\overline{B^{0}_{d}} \rightarrow D^{+} e^{-} {\bar\nu_e}$ [15], the spectrum
of $\overline{B}^0 \rightarrow D^{+*} e^{-} {\bar\nu_e}$ [3] and the ratio
${\Gamma(B^{-} \rightarrow D^{0*}_{L} e^{-} {\bar\nu_e }) /
 \Gamma(B^{-} \rightarrow D^{0*}_{T} e^{-} {\bar\nu_e})}$ [16]
we obtain the values for the free parameters
of the chosen expressions for $\xi(w^2)$:
\begin{eqnarray}
\xi^{par}(w^2) & \equiv & 1+b~w^2+c~w^4
\label{15bis} \\
\xi^{exp}(w^2) & \equiv & \exp\left\{\beta w^2\right\}
\label{15ter}\\
\xi^{pole}(w^2) & \equiv &  \left(1-\frac{w^2}{w_0^2}\right)^{-1}
\label{15quater}\\
\xi^{phys.}_{pole}(w^2) & \equiv &
\xi(0) \left(1-\frac{w^2}{1.69^2} \right)^{-1}
\label{15penta}
\end{eqnarray}
The values for the annihilation parameters
$W_{PP}$, $W_{DV}$ and $W_{D^{*}P}$ are obtained from the
experimental values of the rates
$\bar B^0_d\rightarrow D^+\pi^-$, $\bar B^0_d\rightarrow D^+\rho^-$ and
$\bar B^0_d\rightarrow D^{+*}\pi^-$ respectively and are consistent with the
experimental upper limits on $\bar B^0_d\rightarrow D^{+}_{s}K^{-}$ and
$\bar B^0_d\rightarrow D^{0}\rho^{0}$.
These results are shown in Table 2, where also we report the predictions
for the $W$'s following Fakirov and Stech approach [17],[10].

In Table 3, in correspondence to the values in Table 2,
we compare  the theoretical predictions and the
experimental values for the nonleptonic widths of $B$ mesons decays.
We put in the same box the  rates which should be equal
according to the
$\Delta I = 0 $ selection rule, obeyed by the Cabibbo favoured part of
the $\Delta B=\Delta S=1$ hamiltonian. The amplitudes for the decay
channels $B^{-},\overline{B}^{0}_{d} \rightarrow D \rho$ are
constrained by the $\Delta I =1$, $\Delta I_{3} = -1$ selection rule
to obey:
\begin{equation}
{\cal A}(\overline{B}^{0}_{d} \rightarrow D^{+} \rho^{-})
-{\cal A}(B^{-} \rightarrow D^{0} \rho^{-}) = \sqrt{2} ~
{\cal A}(\overline{B}^{0}_{d} \rightarrow D^{0} \rho^{0})
\label{16}
\end{equation}
which is consistent with the data.

{}From Tables 2 and  3, one sees that the parabolic form (\ref{15bis})
is slightly favoured with
respect to the others. In the case of the physical pole expression for
$\xi(w^2)$ (\ref{15penta}) we find $\xi(0)=0.84\pm 0.05$ three standard
deviations away from the value 1 predicted by
$CVC$ and $CAC$ symmetries of $HQET$, for which one expects breaking
effects only of the order $\Lambda_{QCD}^{2}/m_{c}^{2}$ and
$\alpha_{s}(m_{c})/\pi$.
The central values for the $W$'s come larger in magnitude than the
predictions in [10], as it has been the case for charm decays, but in
general they are consistent with a vanishing value with the only exception
of $W_{PP}$ found in correspondence to the parabolic parameterization
for $\xi(w^2)$.
The amplitude for $B\rightarrow J/\Psi~K$,
which does not depend on $\xi(w^2)$, is proportional to
$f_{+}^{(bq)}(m^{2}_{J/\Psi})$, which by $SU(2)_{HF}$ is given by
(\ref{9}) and (\ref{11}) in terms of the $|\Delta C|=1$ form factors
studied by $E691$.\\
The agreement of the predictions with the measured rates represents
a positive test for  $SU(2)_{HF}$.
\bigskip\bigskip
\par\noindent
{\bf 3. ${\bf CP}$ violation in ${\bf \Delta C=1}$ neutral beauty decays}
\bigskip
\par
In previous papers [10] we studied the $CP$ violating asymmetries in the
decays of neutral beautiful particles in a different theoretical
framework, consisting in a pole behaviour for all the $\Delta B=1$
form factors, $SU(4)$ flavour symmetry and in the assumption that the
$\Delta B=1$ vector and
axial charges at $p_{z} = \infty$ are renormalized by the factor $0.8$,
derived by the semileptonic decay of $D$ mesons. It is therefore interesting
to give the predictions within the theoretical framework assumed here.
We can give definite predictions for the $CP$ violating asymmetries
only for $|C|=1$ final states with the annihilation contributions taken
from the data on Cabibbo favoured decays.

The mixing in the $\overline{B_{d}}^{0}-B_{d}^{0}$, which depends on
the parameter
$z_{d}=\Delta M_{d}/\Gamma_{d}=0.72 \pm 0.14~$[18], gives rise, in presence of
$CP$ violation, to non-vanishing values for the asymmetries:
\begin{equation}
C^{d}_{f} = \frac{\int_{0}^{\infty} dt \left\{ \Gamma\left[B^{0}_{d}(t)
\rightarrow f\right] -\Gamma\left[\overline{B_{d}}^{0}(t)
\rightarrow \bar{f}\right]\right\} }
{\int_{0}^{\infty} dt \left\{ \Gamma\left[B^{0}_{d}(t)
\rightarrow f\right] +\Gamma\left[\overline{B_{d}}^{0}(t)
\rightarrow \bar{f}\right]\right\}}
\label{17}
\end{equation}
where $\Gamma\left[B^{0}_{d}(t) \rightarrow f\right]$ is the rate
for producing at the instant $t$ the final state $f$ from a  state, which
at $t=0$ is at $B^{0}_{d}$ and $\bar{f} = (CP) f$.

One has [19] :
\begin{equation}
C^{d}_{f} = \frac{2 z_{d}}{2 + z_{d}^{2} + z_{d}^{2} \left| x^{d}_{f}
\right|^2}
Im\left[ \frac{ V_{bt} V^{*}_{dt}}{V_{bt}^{*} V_{dt}} x^{d}_{f} \right]
\label{18}
\end{equation}
with:
\begin{equation}
x^{d}_{f} = \frac{ {\cal A}\left( B^{0}_{d} \rightarrow \bar{f} \right) }
{ {\cal A}\left( \overline{B^{0}}_{d} \rightarrow \bar{f} \right) }
= \frac{ {\cal A}^{*}\left( \overline{B^{0}}_{d} \rightarrow f \right) }
{ {\cal A}\left( \overline{B^{0}}_{d} \rightarrow \bar{f} \right) }
\label{19}
\end{equation}
For the asymmetries associated to $B_{s}^{0}$ one can simply substitute
$d \rightarrow s$ in (\ref{18}) and ({\ref{19}).\\
The $x^{q}_{f}$ defined in (\ref{19}) ($q=d,s$) is given by the product of
two terms:
\begin{equation}
x^{q}_{f} = S_{f}^{q} R^{q}_{f}
\label{19bis}
\end{equation}
where $S_{f}^{q}$ is a function of the only CKM matrix
elements and $R^{q}_{f}$ depends on the explicit computation of the
matrix elements of the weak effective hamiltonian. By substituting
(\ref{19bis}) in  (\ref{18}) one obtains, neglecting the final state
interaction:
\begin{equation}
C^{q}_{f} = \frac{2 z_{q}R^{q}_{f}}{2 + z_{q}^{2} + z_{q}^{2} (R^{q}_{f})^2
\left| S_{f}^{q} \right|^2}
Im\left[ \frac{ V_{bt} V^{*}_{qt}}{V_{bt}^{*} V_{qt}} S_{f}^{q} \right].
\label{20}
\end{equation}
For all the final states with $|C|=1$
$S_{{\overline f}}^{q}=1/S_{ f}^{q}$ and for all $f$ with
$C=1$ we have
\footnote
{We follow the Kobayashi-Maskawa parameterization reported also in [10]}:
\begin{equation}
\left|S_{f}^{s}\right|^{2} = s_1^4\left| S_{f}^{d}\right|^{2}\cong
 \frac{s_{2}^{2} +s_{3}^{2} + 2 s_{2} s_{3}c_{\delta}}{s_{3}^{2}}
\label{21}
\end{equation}
In the Table 4 and 5 we report the expressions of $R^{q}_{f}$ found
within the previous [6] and present approach, and  the approximate expressions
of $Im\left[ \frac{ V_{bt} V^{*}_{qt}}{V_{bt}^{*} V_{qt}} S_{f}^{q} \right]$
for the decay channels with $|C|=1$ and with $\pi$, $\rho$, $K$ or
$K^*$ in the final state. As in [8] we define
\begin{equation}
w^{(*)2}_{\pi}=\frac{m_{\pi}^2-(m_B-m_{D^{(*)}})^2}{m_Bm_{D^{(*)}}}~.
\label{22bis}
\end{equation}
In Table 6 we report the values of $\xi(w^2)$ which appear in
$R^{q}_{f}$.\\
As one can see in Table 4 for the decays with neutral particles in
the final state, the central values of the annihilation contributions
are of the same order of magnitude than the direct ones, which are
proportional to $k_2 \cong -0.25$ according to equations (\ref{3}),
and their relative sign changes, when we change the initial (or the
final state) into its $CP$ conjugate. In particular for
$D^0(D^{*0})\pi^0$ and $D^0\rho^0~$ final states the Cabibbo
allowed amplitudes are expected to be small and with uncertain sign in
such a way to allow for
values of $|x^d_f|=|R^d_fS^d_f|\cong 1$ and consequently for large
asymmetries of either sign.

Instead for the final states with charged particles we are able to
predict the sign and the order of magnitude of the corresponding
$R^q_f$. The predictions for $C^d_f$ and $C^s_f$ with the values of the
Kobayashi-Maskawa parameters
\begin{equation}
\begin{array}{cccc}
s_2 = 0.050  & s_3 = 0.023 & s_{\delta} = 0.894 & c_{\delta}  = -0.448\\
\end{array}
\label{23}
\end{equation}
are reported in Tables 7 and 8 respectively. In correspondence of
$\xi^{par}(w^2)$ the branching ratios and the minimum number of
$b\bar b$-couples needed for a 3 $\sigma$-evidence of the asymmetries are also
computed.

\bigskip\bigskip
\par\noindent
{\bf 5. Conclusion}
\bigskip
\par
\par\noindent
The measured branching ratios for two body nonleptonic decays of B-particles
are well described, in absence of final state interaction,
by assuming factorization,
total colour screening and with the matrix elements of the weak currents
given by Isgur and Wise theory and $SU(2)_{HF}$.
We allow for $SU(3)$ invariant contributions of the annihilation terms
to be fitted from data, which come out larger, as in charm decay [4],
than the values predicted by assuming the dominance of the lowest $0^{\pm}$
octets [17]. Between the
parameterization considered for the universal function $\xi(w^{2})$ the
parabolic one is slightly favoured, expecially for the spectrum in the
$e~\bar{\nu}$ invariant mass for the decay
$\overline{B^0} \rightarrow D^{+*} e^{-}\bar{\nu}_{e}$.
This parameterization is also the one giving $|V_{bc}|$
the value  $0.044$ in best agreement with present determinations.\\
Indeed, also with a $\xi(w^{2})$ with the pole corresponding to the $b\bar{c}$
resonances and $\xi(0) \neq 1$ one gets a good fit, just showing to us that
the large errors in the data cannot at the moment allow for drastic
conclusions. Instead
within the previous approach based on $SU(4)_{(u,d,s,c)}$ one predicts for
the rate $B \rightarrow J/\Psi K$ larger values than experiment, which
casts doubts on its validity  despite its capability in describing the
rates into final states with $|C|=1$. With the
parameters fixed from the measured decays we are able to predict the
$CP$ violating asymmetries in the decays of neutral $B$-particles into
$|C|=1$ final states in absence of final state interaction. Large
asymmetries are expected for the final states $D^{(*)0}\pi^0$ and
$D^{0}\rho^0$, but the present uncertainties do not
allow to predict even the sign.\\
We have better confidence  in our predictions on sign and order of
magnitude of the asymmetries for the decays with charged particles in
the final state.

\newpage
{\bf References}
\bigskip
\begin{itemize}
\itemindent=-1cm
\item[[1]]N. Isgur and M.B. Wise Phys. Lett.  {\bf B232}(1989)113;
\item[]$~$N. Isgur and M.B. Wise Phys. Lett.  {\bf B237}(1990)527;
\item[[2]]M. Neubert, Phys. Lett. {\bf B264}(1991)455 ;
\item[[3]]Argus Collaboration data presented by H. Scroder
 to Desy Workshop (Hamburg 1992);
\item[[4]]F. Buccella,M.Lusignoli, G. Miele and A. Pugliese, Zeit. Phys. C
{\bf 55}(1992)243;
\item[[5]]J.C. Anjos et al. ($E691$ Collaboration), Phys. Rev. Lett.
{\bf 62}(1989)1587 and {\bf 65}(1990)2630;
\item[[6]]N. Isgur and M.B. Wise, Phys. Rev. {\bf D42}(1990)2388;
\item[[7]]M. Tanimoto, Phys. Rev. {\bf D44}(1991)1449;
\item[[8]]F. Buccella,F. Lombardi and P.Santorelli, Il Nuovo Cimento A,
{\bf 105}(1992)993,
\item[~~~] G.Burdman, Phys. Lett. {\bf B284}(1992)133;
\item[[9]]J. Rosner, Phys. Rev. {\bf D$42$}(1990),3732;
\item[$~~~$]T.Mannel, W.Roberts and Z.Rizak, Phys. Lett.
{\bf B$259$}(1991),359;
\item[[10]]F. Buccella,G.Mangano and G. Miele, Il Nuovo Cimento A,
{\bf 104}(1991)1293;
\item[~~]F. Buccella,G.Mangano,G. Miele and P.Santorelli, Il Nuovo Cimento A
{\bf 105}(1992)33;
\item[[11]]A.F.Falk, H.Georgi, B.Grinstein and M.B.Wise, Nucl. Phys.
{\bf B343}(1990)1;
\item[[12]]G.Altarelli, G.Curci, G.Martinelli and S.Petrarca, Nucl. Phys.
{\bf B187}(1981)461;
\item[[13]]V. Lubicz, G. Martinelli and C.T. Sachrajda, Rome Preprint
n. 748, July 11th, 1990;
\item[[14]]M. Bauer and B. Stech, Phys. Lett. {\bf B152}(1985)380;
\item[~~]M. Bauer, B. Stech and Wirbel, Zeit. fur Phys. {\bf C34}(1987)103;
\item[[15]]Particle Data Group, Phys. Rev. {\bf D45}(1992)Part 2;
\item[[16]]J.G. Korner, K.Schilcher, M.Wirbel and J.L.Wu, Zeit. Phys. C
{\bf 48}(1990)663;
\item[[17]]D.Fakirov and B.Stech, Nucl. Phys. {\bf B133}(1978)315;
\item[[18]]H.Albrecht et al.,Phys. Lett.{\bf B192}(1987)245;
\item[[19]]A.B.Carter and A.I.Sanda,Phys.Rev.Lett.{\bf 45}(1980)952,
Phys. Rev.{\bf D23}(1981)1567;
\item[~~~]A.I.Bigi and A.I.Sanda,Nucl.Phys.{\bf B193}(1981)85.
\end{itemize}
\newpage
\par\noindent
{\bf Table 1 }\\
\begin{center}
\begin{tabular}{|c|c|c|c|}
\hline
& & & \\
Form & &$M_{c\bar{q}}(J^{P})$ & $M_{c\bar{q}}(J^{P})$ \\
& Residua & $\{ \bar{q}=\bar{s}\}$& $\{\bar{q}=\bar{u},\bar{d}\}$\\
Factors & & $GeV$ & $GeV$\\
& & & \\
\hline
& & & \\
$f_{0}^{(cq)}(q^{2})$ & $0.79 \pm 0.08 $&  $2.60~(0^{+})$ & $2.47~(0^{+})$ \\
& & & \\
$f_{1}^{(cq)}(q^{2})$ & $0.79 \pm 0.08 $& $2.11~(1^{-})$ & $2.01~(1^{-})$ \\
& & & \\
$A_{0}^{(cq)}(q^{2})$ & $0.71 \pm 0.16 $& $1.97~(0^{-})$ & $1.87~(0^{-})$ \\
& & & \\
$A_{1}^{(cq)}(q^{2})$ & $0.46 \pm 0.07 $& $2.53~(1^{+})$ & $2.42~(1^{+})$ \\
& & & \\
$A_{2}^{(cq)}(q^{2})$ & $0.00 \pm 0.22 $& $2.53~(1^{+})$ & $2.42~(1^{+})$ \\
& & & \\
$V^{(cq)}(q^{2})$    &  $0.90 \pm 0.32 $& $2.11~(1^{-})$ & $2.01~(1^{-})$ \\
& & & \\
\hline
\end{tabular}
\end{center}
\newpage
\par\noindent
{\bf Table 2 }\\
\begin{center}
\begin{tabular}{|c|c|c|c|c|}
\hline
& & & &\\
& $\xi^{par}(w^{2})$ & $\xi^{exp}(w^{2})$ & $\xi^{pole}(w^{2})$
& $\xi^{phys.}_{pole}(w^{2})$ \\
& & & &\\
\cline{2-5}
& & & &\\
& $\left\{ \begin{array}{c} b=1.06^{+0.20}_{-0.23} \\ c=0.62^{+0.17}_{-0.20}
\end{array}\right.$ & $\beta=0.48^{+0.15}_{-0.14} $ & $w_{0}=1.20
^{+0.27}_{-0.17}$
& $\xi(0)=0.83^{+0.14}_{-0.09}$\\
& & & &\\
\hline
& & & &\\
$|V_{bc}|$ & $0.044\pm 0.004$ & $0.037\pm 0.003$
& $0.039\pm 0.004$ & $0.040\pm 0.005$\\
& & & &\\
\hline
& & & &\\
$W_{PP}$ & $-0.078^{+0.019}_{-0.021}$ & $-0.028^{+0.022}_{-0.025}$
& $-0.032^{+0.021}_{-0.024}$ & $-0.051^{+0.021}_{-0.023}$\\
& & & &\\
\hline
& & & &\\
$W_{DV}$ & $+0.115^{+0.105}_{-0.152} $ & $+0.047^{+0.125}_{-0.181}$  &
 $+0.051^{+0.119}_{-0.172} $ & $+0.085^{+0.116}_{-0.168}$ \\
& & & &\\
\hline
& & & &\\
$W_{D^{*}P}$ & $-0.054^{+0.022}_{-0.025}$ & $-0.042^{+0.026}_{-0.029}$ &
$ -0.042^{+0.025}_{-0.028} $ &$-0.058^{+0.024}_{-0.027}$ \\
& & & &\\
\hline
& & & &\\
$\chi^2$ & $13.35 $ & $18.87$ & $17.83 $ & $18.71$\\
$(\chi^2/Ndf)$ & $(0.89)$ & $(1.18)$ & $(1.11)$ & $(1.17)$ \\
& & & &\\
\hline
& & & &\\
$\chi^{2}_{spectrum}$ & $6.79$ & $8.01$ & $7.54$ & $11.05$ \\
& & & &\\
\hline
\end{tabular}\\
\footnotesize{The order of magnitude of the analogous of $W$'s estimated in
[17] and [10] are $W_{PP}\cong -0.002$,
$W_{DV}=W_{D^{*}P}\cong -0.010$ }
\end{center}
\newpage
\par\noindent
{\bf Table 3 }\\
\footnotesize
\begin{center}
\begin{tabular}{|l|c|c|c|c|c|}
\hline
& & \multicolumn{4}{c|}{} \\
     & Exp. rates [15]  & \multicolumn{4}{c|}{ Theoretical rates } \\
& & \multicolumn{4}{c|}{} \\
\cline{2-6}
& & & & &\\
$~~~$Channels  &   &  $\xi^{par}(w^2)$      &  $\xi^{exp}(w^2)$
&  $\xi^{pole}(w^2)$ &  $\xi^{phys.}_{pole}(w^2)$  \\
& & & & &\\
& $(10^{-12}~MeV)$ & $(10^{-12}~MeV)$  & $(10^{-12}~MeV)$  & $(10^{-12}~MeV)$
& $(10^{-12}~MeV)$ \\
& & & & &\\
\hline
& & & & &\\
$\begin{array}{l} B^{-} \rightarrow D^{0} D_{s}^{-}\\
\overline{B^{0}_{d}} \rightarrow D^{+} D_{s}^{-} \end{array}$ & $\left.
\begin{array}{c} 9.1 \pm 5.3\\ 3.7 \pm 2.3 \end{array} \right\}$ &
$4.2$ & $4.7$ & $4.6$ & $4.9$\\
& & & & &\\
\hline
& & & & &\\
$B^{-} \rightarrow D^{0} \pi^{-}$ & $1.81 \pm 0.53$ &
$1.81$ & $1.06$ & $1.08$ & $1.34$\\
& & & & &\\
\hline
& & & & &\\
$\overline{B^{0}_{d}} \rightarrow D^{+} \pi^{-}$ & $1.48 \pm 0.32$
& $1.48$ & $1.48$ & $1.48$ & $1.48$\\
& & & & &\\
\hline
& & & & &\\
$\overline{B^{0}_{d}} \rightarrow D_{s}^{+} K^{-}$ & $ < 0.62$
& $0.27$ & $0.03$ & $0.04$& $0.09$\\
& & & & &\\
\hline
& & & & &\\
$ \begin{array}{l} B^{-} \rightarrow J/\Psi K^{-}\\
\overline{B^{0}_{d}} \rightarrow J/\Psi \overline{K^{0}} \end{array}$ & $\left.
\begin{array}{c} 0.37 \pm 0.10\\ 0.30 \pm 0.14 \end{array} \right\}$ &
$0.38$ & $0.27$ & $0.30$ & $0.31$\\
& & & & &\\
\hline
& & & & &\\
$\overline{B^{0}_{d}} \rightarrow D^{+*} D_{s}^{-}$ & $7.4 \pm 5.1$
& $2.83$ & $3.29$ & $3.24$ & $3.23$\\
& & & & &\\
\hline
& & & & &\\
$B^{-} \rightarrow D^{0} \rho^{-}$ & $6.2 \pm 2.9$
& $4.83$ & $3.36$ & $3.40$& $4.07$\\
& & & & &\\
\hline
& & & & &\\
$\overline{B^{0}_{d}} \rightarrow D^{+} \rho^{-}$ & $4.2 \pm 2.8$
& $4.17$ & $4.17$ & $4.17$& $4.17$\\
& & & & &\\
\hline
& & & & &\\
$\overline{B^{0}_{d}} \rightarrow D^{0} \rho^{0}$ & $< 0.28$
& $0.010$ & $0.021$ & $0.019$& $0.$\\
& & & & &\\
\hline
& & & & &\\
$B^{-} \rightarrow D^{0*} \pi^{-}$ & $2.5 \pm 0.7$
& $1.31$ & $1.20$ & $1.18$& $1.37$\\
& & & & &\\
\hline
& & & & &\\
$\overline{B^{0}_{d}} \rightarrow D^{+*} \pi^{-}$ & $1.48 \pm 0.32$
& $1.48$ & $1.48$ & $1.48$ & $1.48$\\
& & & & &\\
\hline
\end{tabular}
\end{center}
\newpage
\normalsize
\par\noindent
{\bf Table 4 }\\
\large
\begin{center}
\begin{tabular}{|c|c|c|c|}
\hline
 & & & \\
$f$ &$\begin{array}{c}
R^{d}_{f} \\ cfr.~[10]
\end{array}$ & $ R^{d}_{f}$ & $Im\left[ \frac{V_{bt}V_{dt}^{*}}
{V_{bt}^{*}V_{dt}}S^{d}_{f}\right]\cong$\\
 & &  & \\
\hline
 & &  & \\
$D^{+} \pi^{-}$  &$0.49  $&$
\left(\frac{0.43\xi(w^2_{\pi})+W_{PP})}{0.27-W_{PP}}\right)$ &
$ -\frac{s_{\delta}( s_{2}+2s_{3} c_{\delta})}{ s_{1}^{2} s_{3}}$\\
 & &  & \\
$D^{-} \pi^{+}$ &$2.06 $&$
\left(\frac{0.27-W_{PP}}{0.43\xi(w^2_{\pi})+W_{PP}}\right)$ &
$ -\frac{s_{1}^{2} s_{3} s_{\delta} ( s_{2}+2s_{3} c_{\delta})}
{s_{3}^{2} +s_{2}^{2} + 2 s_{2}s_{3} c_{\delta}}  $\\
 & &  & \\
$D^{0} \pi^{0}$ &$1.04  $&$
\left(\frac{0.06+W_{PP}}{0.06-W_{PP}}\right)$ & $-
\frac{s_{\delta}( s_{2}+2s_{3} c_{\delta})}{ s_{1}^{2} s_{3}} $\\
 & &  & \\
$\overline{D}^{0} \pi^{0}$ &$0.96 $&$
\left(\frac{0.06-W_{PP}}{0.06+W_{PP}}\right)$ & $
-\frac{s_{1}^{2} s_{3} s_{\delta} ( s_{2}+2s_{3} c_{\delta})}
{s_{3}^{2} +s_{2}^{2} + 2 s_{2}s_{3} c_{\delta}} $\\
 & &  & \\
$D^{+}_{s} K^{-}$ &$ -1  $& $ -1$ & $\frac{s_{2}s_{\delta}}{s_{3}} $\\
 & &  & \\
$D^{-}_{s} K^{+}$ &$ -1  $& $-1$  & $\frac{s_{2} s_{3} s_{\delta}}
{s_{3}^{2} +s_{2}^{2} + 2 s_{2}s_{3} c_{\delta}} $\\
 & &  & \\
$D^{0*} \pi^{0}$ &$1.17$&$
\left(\frac{0.07+W_{D^{*}P}}{0.07-W_{D^{*}P}}\right)$ & $
\frac{s_{\delta}( s_{2}+2s_{3} c_{\delta})}{ s_{1}^{2} s_{3}} $\\
 & &  & \\
$\overline{D}^{0*} \pi^{0}$&$0.85  $& $
\left(\frac{0.07-W_{D^{*}P}}{0.07+W_{D^{*}P}}\right)$ & $
\frac{s_{1}^{2} s_{3} s_{\delta} ( s_{2}+2s_{3} c_{\delta})}
{s_{3}^{2} +s_{2}^{2} + 2 s_{2}s_{3} c_{\delta}} $\\
  & & & \\
$D^{0} \rho^{0}$ &$1.18 $& $
\left(\frac{0.09-W_{DV}}{0.09+W_{DV}}\right)$  & $
\frac{s_{\delta}( s_{2}+2s_{3} c_{\delta})}{ s_{1}^{2} s_{3}} $\\
 & &  & \\
$\overline{D}^{0} \rho^{0}$ &$0.85$&$
\left(\frac{0.09+W_{DV}}{0.09-W_{DV}}\right)$ & $
\frac{s_{1}^{2} s_{3} s_{\delta} ( s_{2}+2s_{3} c_{\delta})}
{s_{3}^{2} +s_{2}^{2} + 2 s_{2}s_{3} c_{\delta}}$\\
 & &  & \\
$D^{+} \rho^{-}$ &$0.97$&$
\left(\frac{0.81\xi(w^2_{\rho})-W_{DV}}{0.40+W_{DV}}\right)$ & $
\frac{s_{\delta}( s_{2}+2s_{3} c_{\delta})}{ s_{1}^{2} s_{3}} $\\
 & &  & \\
$D^{-} \rho^{+}$ &$1.03$&
$\left(\frac{0.40+W_{DV}}{0.81\xi(w^2_{\rho})-W_{DV}}\right)$  & $
\frac{s_{1}^{2} s_{3} s_{\delta} ( s_{2}+2s_{3} c_{\delta})}
{s_{3}^{2} +s_{2}^{2} + 2 s_{2}s_{3} c_{\delta}}$\\
 & &  & \\
$D^{+*} \pi^{-}$ &$0.51  $&
$\left(\frac{0.48\xi(w^{*2}_{\pi})+W_{D^{*}P}}{0.30-W_{D^{*}P}}\right)$ & $
\frac{s_{\delta}( s_{2}+2s_{3} c_{\delta})}{ s_{1}^{2} s_{3}} $\\
 & &  & \\
$D^{-*} \pi^{+}$ & $1.95$&$
\left(\frac{0.30-W_{D^{*}P}}{0.48\xi(w^{*2}_{\pi})+W_{D^{*}P}}\right)$ & $
\frac{s_{1}^{2} s_{3} s_{\delta} ( s_{2}+2s_{3} c_{\delta})}
{s_{3}^{2} +s_{2}^{2} + 2 s_{2}s_{3} c_{\delta}}$\\
 & &  & \\
\hline
\end{tabular}
\end{center}
\newpage
\par\noindent
\normalsize
{\bf Table 5 }\\
\begin{center}
\begin{tabular}{|c|c|c|c|}
\hline
 & &  & \\
$f$ & $\begin{array}{c}
R^{s}_{f}\\ cfr.~[10] \end{array}$
& $R^{s}_{f}$ & $Im\left[ \frac{V_{bt}V_{st}^{*}}
{V_{bt}^{*}V_{st}}S^{s}_{f}\right]\cong$\\
 & &  & \\
\hline
 & &  & \\
$D^{+}_{s} K^{-}$ &$1.59 $&$
\left(\frac{0.54\xi(w^2_{K})+W_{PP}}{0.26-W_{PP}}\right)$ & $
\frac{s_{2}s_{\delta} }{s_{3}}$\\
 & &  & \\
$D^{-}_{s} K^{+}$ &$0.63$&$
\left(\frac{0.26-W_{PP}}{0.54\xi(w^2_{K})+W_{PP}}\right)$ & $
\frac{s_{2}s_{3}s_{\delta}}{s_{2}^{2}+s_{3}^{2}+ 2 s_{2} s_{3} c_{\delta}}$\\
 & &  & \\
$D^{+*}_{s} K^{-}$ &$0.66$&$
\left(\frac{0.61\xi(w^2_{K})+W_{D^{*}P}}{0.29-W_{D^{*}P}}\right)$ &
$-\frac{s_{2}s_{\delta} }{s_{3}}$\\
 & &  & \\
${D}^{-*}_{s} K^{+}$ &$1.52$&$
\left(\frac{0.54\xi(w^2_{K})+W_{PP}}{0.26-W_{PP}}\right)$ & $
-\frac{s_{2}s_{3}s_{\delta}}{s_{2}^{2}+s_{3}^{2}+ 2 s_{2} s_{3} c_{\delta}}$\\
 & &  & \\
$D^{+}_{s} K^{-*}$ &$0.73$&$
\left(\frac{0.39\xi(w^2_{K^*})-W_{DV}}{0.30+W_{DV}}\right)$ &
$ - \frac{s_{2}s_{\delta} }{s_{3}}$\\
 & &  & \\
$D^{-}_{s} K^{+*}$ &$1.38$&$
\left(\frac{0.30+W_{DV}}{0.39\xi(w^2_{K^*})-W_{DV}}\right)$ & $
-\frac{s_{2}s_{3}s_{\delta}}{s_{2}^{2}+s_{3}^{2}+ 2 s_{2} s_{3} c_{\delta}}$\\
 & &  & \\
\hline
\end{tabular}
\end{center}
\normalsize
\par\noindent
{\bf Table 6  }\\
\begin{center}
\begin{tabular}{|c|c|c|c|c|}
\hline
& & & & \\
$~~$  & $\xi^{par}(w^2)$ & $\xi^{exp}(w^2)$& $\xi^{pole}(w^2)$ &
$\xi^{phys.}_{pole}(w^2)$\\
& & & & \\
\hline
& & & & \\
$w^2_{\pi}$ & $0.611$ &$0.569$ &$0.550$ &$0.588$ \\
& & & & \\
\hline
& & & & \\
$w^2_{\rho}$ & $0.590$ &$0.585$ &$0.563$ &$0.596$ \\
& & & & \\
\hline
& & & & \\
$w^2_{K}$ & $0.577$ & $0.597$ & $0.572$ & $0.603$ \\
& & & & \\
\hline
& & & & \\
$w^2_{K^*}$ & $0.565$ & $0.612$ &  $0.584$ & $0.611$ \\
& & & & \\
\hline
& & & & \\
$w^{*2}_{\pi}$ & $0.561$ & $0.617$ & $0.589$ & $0.614$ \\
& & & & \\
\hline
\end{tabular}\\
\end{center}
\newpage
\normalsize
\par\noindent
{\bf Table 7 }\\
{\bf \footnotesize
\begin{center}
\begin{tabular}{|c|c|c|c|c|c|}
\hline
 & & & & & \\
$ C_{f}^{d}$ & $\xi^{par}(w^2)$ & $\xi^{exp}(w^2)$ & $\xi^{pole}(w^2)$ &
 $Br(B^{0}_{d},f_{i})$ & $\sigma(B^{0}_{d} B^{-}) \epsilon N_{b\bar{b}}$\\
 & & & & & \\
 $f$ & & & & $\%$ & \\
 & & & & & \\
\hline
 & & & & & \\
$D^{+} \rho^{-}$ &$0.059$ &$0.043$ &$0.046$ &$0.12~10^{-2}$ &$8.0~10^{5}$ \\
 & & & & &  \\
\hline
 & & & & & \\
$D^{+} \pi^{-}$ & $-0.079$ & $-0.058$ &$-0.062$ &$0.76~10^{-3}$ &
$1.2~10^{6}$ \\
 & & & & &  \\
\hline
 & & & & & \\
$D^{+*} \pi^{-}$ &$0.068$ &$0.056$ &$0.059$ &$0.56~10^{-3}$ &$1.7~10^{6}$ \\
 & & & & &  \\
\hline
 & & & & & \\
$D^{-} \rho^{+}$ &$0.012$ &$0.009$ &$0.010$ &$0.96$ &$3.8~10^{6}$ \\
 & & & & &  \\
\hline
 & & & & & \\
$D^{-} \pi^{+}$ &$-0.016$ &$-0.012$ &$-0.013$ &$0.34$ &$5.9~10^{6}$ \\
 & & & & &  \\
\hline
 & & & & & \\
$D^{-*} \pi^{+}$ &$0.014$ &$0.012$ &$0.012$ &$0.340$ &$8.1~10^{6}$ \\
 & & & & &  \\
\hline
 & & & & & \\
$D^{+}_{s} K^{-}$ &$0.042$ &$0.042$ &$0.042$ &$0.37~10^{-4}$ &$2.5~10^{7}$ \\
 & & & & &  \\
\hline
 & & & & & \\
$D^{-}_{s} K^{+}$ &$0.009$ &$0.009$ &$0.009$ &$0.61~10^{-1}$ &$1.2~10^{8}$ \\
 & & & & &  \\
\hline
\multicolumn{6}{|c|}{}\\
\hline
 & & & & & \\
$D^{0} \rho^{0}$ &$-0.285$ &$0.130$ &$0.146$ &$0.94~10^{-4}$ &$1.2~10^{7}$ \\
 & & & & &  \\
\hline
 & & & & & \\
$D^{0} \pi^{0}$ &$0.284$ &$-0.111$ &$-0.131$ &$0.61~10^{-4}$ &$1.8~10^{7}$ \\
 & & & & &  \\
\hline
 & & & & & \\
$D^{0*} \pi^{0}$ &$0.311$ &$0.174$ &$0.174$ &$0.33~10^{-4}$ &$3.5~10^{7}$ \\
 & & & & &  \\
\hline
 & & & & & \\
$\overline{D}^{0} \rho^{0}$ &$-0.070$ &$0.028$ &$0.031$ &$0.23~10^{-2}$ &
$4.8~10^{7}$ \\
 & & & & &  \\
\hline
 & & & & & \\
$\overline{D}^{0} \pi^{0}$ &$0.069$ &$-0.023$ &$-0.028$ &$0.15~10^{-2}$ &
$7.4~10^{7}$ \\
 & & & & &  \\
\hline
 & & & & & \\
$\overline{D}^{0*} \pi^{0}$ & $0.080$ &$0.038$ &$0.038$ &$0.62~10^{-3}$ &
$1.4~10^{8}$ \\
 & & & & &  \\
\hline
\end{tabular}
\end{center}
\newpage
\par\noindent
\normalsize
{\bf Table 8 }\\
\begin{center}
\begin{tabular}{|c|c|c|c|c|c|}
\hline
 & & & & & \\
$ C_{f}^{s}$ & $\xi^{par}(w^2)$ & $\xi^{exp}(w^2)$ & $\xi^{pole}(w^2)$ &
 $Br(B^{0}_{s},f_{i})$ & $\sigma(B^{0}_{s} B^{-}) \epsilon N_{b\bar{b}}$\\
 & & & & & \\
 $f$ & & & & $\%$ & \\
 & & & & & \\
\hline
 & & & & & \\
$D^{+}_{s} K^{-}$ &$0.079$ &$0.066$ &$0.069$ &$0.15~10^{-1}$ &$3.3~10^{6}$ \\
 & & & & &  \\
\hline
 & & & & & \\
$D^{-}_{s} K^{+}$ &$0.079$ &$0.065$ &$0.068$ &$0.29~10^{-1}$ &$3.3~10^{6}$ \\
 & & & & &  \\
\hline
 & & & & & \\
$D^{-}_{s} K^{+*}$ &$-0.082$ &$-0.067$ &$-0.070$ &$0.20~10^{-1}$ &$3.8~10^{6}$
\\
 & & & & &  \\
\hline
 & & & & & \\
$D^{+}_{s} K^{-*}$ &$-0.082$ &$-0.067$ &$-0.070$ &$0.15~10^{-1}$ &$3.8~10^{6}$
\\
 & & & & &  \\
\hline
 & & & & & \\
$D^{+*}_{s} K^{-}$ &$-0.074$ &$-0.064$ &$-0.067$ &$0.11~10^{-1}$ &$4.1~10^{6}$
\\
 & & & & &  \\
\hline
 & & & & & \\
${D}^{-*}_{s} K^{+}$ &$-0.073$ &$-0.064$ &$-0.066$ &$0.30~10^{-1}$ &$4.1~10^{6}
$ \\
 & & & & &  \\
\hline
\end{tabular}\\
\footnotesize
{To compute these asymmetries we assume $\tau_{B_d}=\tau_{B_s}$ and
 $z_s = z_d \frac{|V_{st}|^2}{|V_{dt}|^2}$.}
\end{center}
\end{document}